\documentclass[sigconf]{acmart}
\usepackage[utf8]{inputenc}
\usepackage[T1]{fontenc}
\usepackage[inline]{enumitem}
\usepackage{diagbox}
\usepackage{booktabs}
\usepackage{amsmath}
\usepackage{amsthm}
\usepackage{amsfonts}
\usepackage{dashbox}
\usepackage{enumitem}
\usepackage{mathbbol}
\usepackage{mathtools}
\usepackage{hyperref}
\usepackage{url}
\usepackage{nicefrac}
\usepackage{microtype}
\usepackage{natbib}
\usepackage{multirow}
\usepackage{graphicx}
\usepackage{subcaption}
\usepackage{etaremune}
\usepackage{xcolor}
\usepackage{bbm}
\usepackage{algorithm}
\usepackage{algorithmic}
\usepackage{tikz}
\usepackage{pgfplots}
\usetikzlibrary{arrows.meta,arrows}
\pgfplotsset{compat=1.14}
\usepackage{balance}
\usepackage{flushend}

\theoremstyle{definition}
\newtheorem{definition}{Definition}

\newcommand{\ie}{\emph{i.e.}}
\newcommand{\eg}{\emph{e.g.}}

\newcommand{\vs}{\emph{vs. }}

\AtBeginDocument{%
  \providecommand\BibTeX{{%
    \normalfont B\kern-0.5em{\scshape i\kern-0.25em b}\kern-0.8em\TeX}}}

\copyrightyear{2021}
\acmYear{2021}
\setcopyright{acmcopyright}\acmConference[SIGIR '21]{Proceedings of the 44th
International ACM SIGIR Conference on Research and Development in Information
Retrieval}{July 11--15, 2021}{Virtual Event, Canada}
\acmBooktitle{Proceedings of the 44th International ACM SIGIR Conference on
Research and Development in Information Retrieval (SIGIR '21), July 11--15, 2021,
Virtual Event, Canada}
\acmPrice{15.00}
\acmDOI{10.1145/3404835.3462804}
\acmISBN{978-1-4503-8037-9/21/07}

\begin{document}

\fancyhead{}

\title[MS MARCO: Benchmarking Ranking Models in the Large-Data Regime]{MS MARCO: Benchmarking Ranking Models\\in the Large-Data Regime}

\author{Nick Craswell}
\email{nickcr@microsoft.com}
\affiliation{%
   \institution{Microsoft}
  \country{USA}
}
\author{Bhaskar Mitra}
\email{bmitra@microsoft.com}
\affiliation{%
   \institution{Microsoft, University College London}
  \country{Canada}
}
\author{Emine Yilmaz}
 \email{emine.yilmaz@ucl.ac.uk}
\affiliation{%
   \institution{University College London}
 \country{United Kingdom}
}
\author{Daniel Campos}
\email{dcampos3@illinois.edu}
\affiliation{%
   \institution{University of Illinois, UC}
\country{USA}
}
\author{Jimmy Lin}
\email{jimmylin@uwaterloo.ca}
\affiliation{%
   \institution{University of Waterloo, Microsoft}
  \country{Canada}
}

\begin{abstract}
    Evaluation efforts such as TREC, CLEF, NTCIR and FIRE, alongside public leaderboard such as MS MARCO, are intended to encourage research and track our progress, addressing big questions in our field. However, the goal is not simply to identify which run is ``best'', achieving the top score. The goal is to move the field forward by developing new robust techniques, that work in many different settings, and are adopted in research and practice. This paper uses the MS MARCO and TREC Deep Learning Track as our case study, comparing it to the case of TREC ad hoc ranking in the 1990s. We show how the design of the evaluation effort can encourage or discourage certain outcomes, and raising questions about internal and external validity of results. We provide some analysis of certain pitfalls, and a statement of best practices for avoiding such pitfalls. We summarize the progress of the effort so far, and describe our desired end state of ``robust usefulness'', along with steps that might be required to get us there.
\end{abstract}



\begin{teaserfigure}
\centering
  \includegraphics[width=0.4\textwidth]{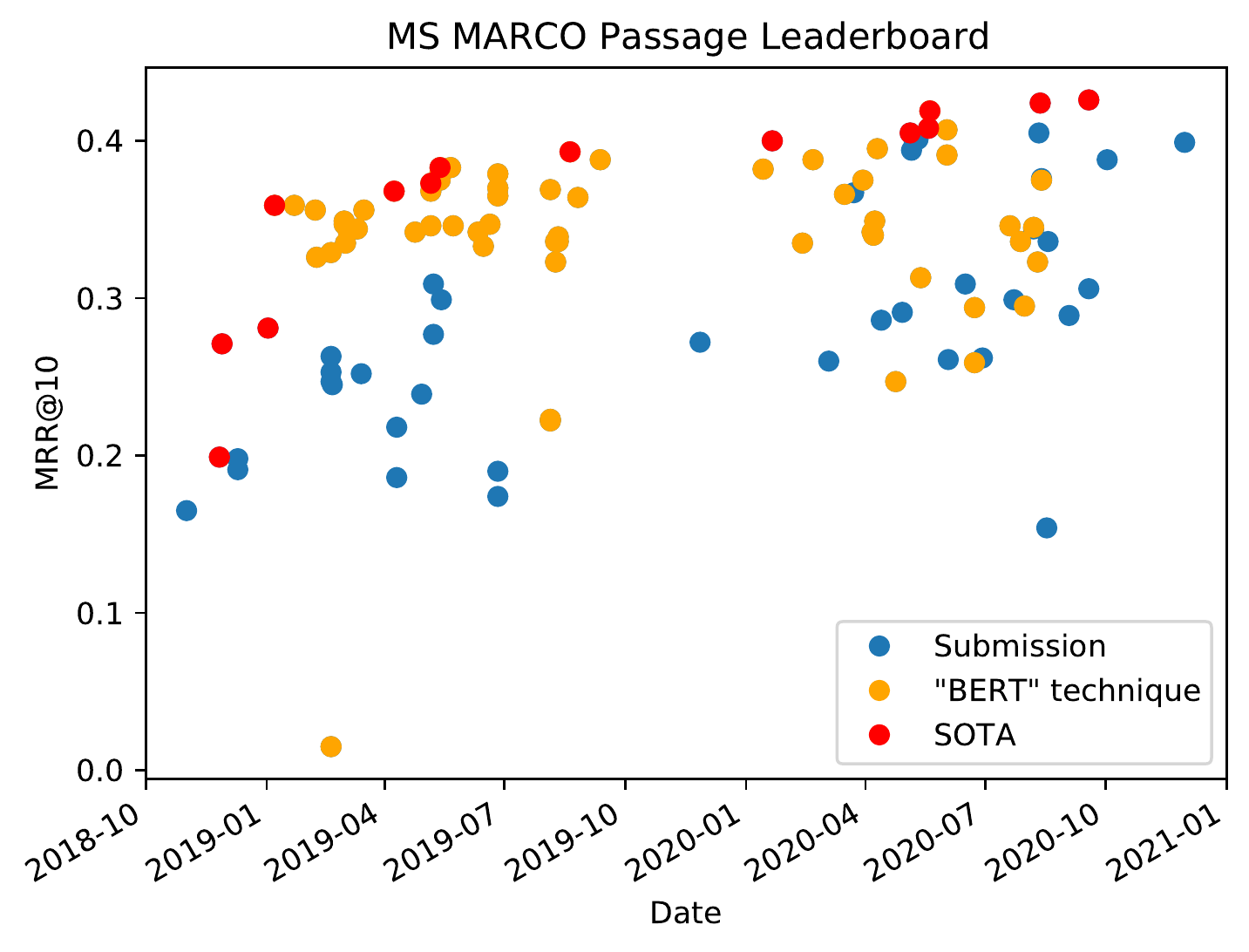}
  \includegraphics[width=0.4\textwidth]{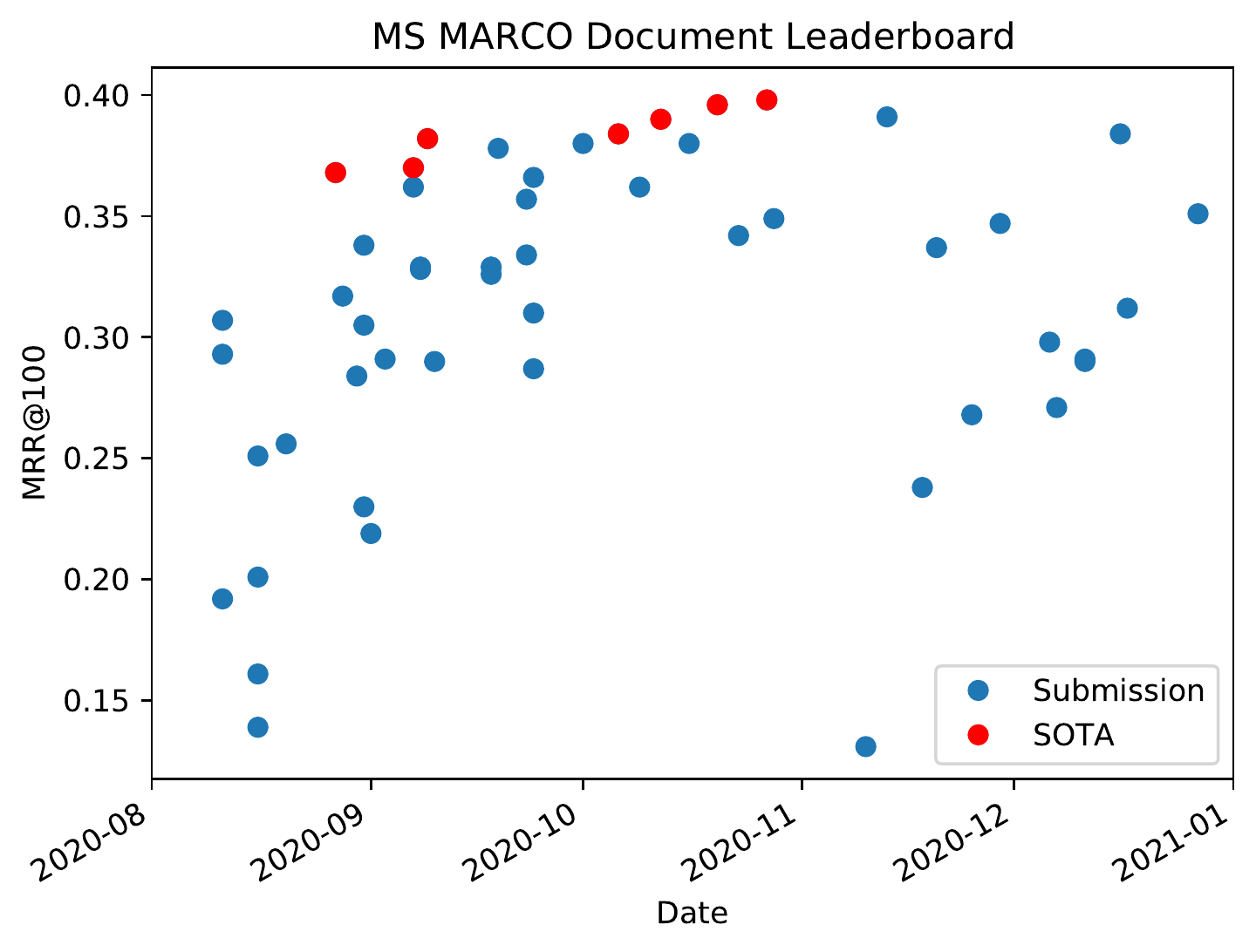}
  \caption{Improvement over time using the MS MARCO passage data (left) and MS MARCO document data (right).}
  \label{fig:teaser}
  \vspace{0.5cm}
\end{teaserfigure}

\maketitle

\section{Introduction}

MS MARCO is a series of datasets, the first of which released in 2016, aiming to help academic researchers explore information access in the large-data regime~\citep{MS_MARCO_ranking}.
The MS MARCO datasets have been a boon for neural IR researchers to support their explorations of ever larger and richer models with an insatiable appetite for more (supervised) training data.
Over the past few years, the datasets have been used in tasks ranging from keyphrase extraction to question answering to text ranking.
Of these tasks, the passage ranking and document ranking tasks have received the most attention from the research community; both are associated with competitive leaderboards\footnote{\url{http://msmarco.org}} and the TREC Deep Learning Track \citep{craswell2021trec, trec2019overview,trec2020overview}. They are standard ad hoc retrieval tasks, with the major difference being the length of the documents that are retrieved:\ the passage ranking task works with paragraph-length segments of text, while the document ranking task works with full-length web pages.

Figure~\ref{fig:teaser} summarizes both leaderboards, passage on the left and document on the right.
The {\it x}-axes represent time, from the introduction of the leaderboards until early 2021.
Each point represents a submission:\ the {\it x}-axis plots the date of submission and the {\it y}-axis plots the official metric (MRR@10 for passage and MRR@100 for document).
Circles in red represent the (current and former) state of the art (SOTA) runs, i.e., a top-scoring run on the leaderboard, beginning with the first submission that beat organizer-supplied baselines.
On left panel in Figure~\ref{fig:teaser} for the passage leaderboard, the large jump in the SOTA in January 2019 represents the work of~\citet{nogueira2019passage}, which is the first known application of pretrained transformers to a ranking task.
This is considered by many to be a watershed moment in IR, as it ushered in a new era of research dominated by the use of pretrained transformer models.
Runs whose description contain the word ``BERT'' are shown in orange in the left panel.
From the multitude of the orange points, we can see the immediate dominance of BERT-based techniques right after its introduction; this is likely even an under-estimate, since there are many ranking models based on pretrained transformer models that do not have BERT in its name (e.g., ELECTRA, T5, etc.).
We did not repeat the same coloring in the document leaderboard because, based on our observations, BERT has become so ingrained that its name is nowadays omitted from the model descriptions.

Prior to the advent of the MS MARCO, deep neural methods in IR were largely being benchmarked on proprietary datasets (\eg, \citep{huang2013learning, mitra2017learning, zamani2018neural}), non-English datasets (\eg, \citep{xiong2017end, dai2018convolutional}), synthetic datasets (\eg, \citep{nanni2017benchmark, van2016learning}), or under weak supervision settings (\eg, \citep{dehghani2017neural, zamani2018neural}).
This made it difficult for the community to compare these emerging methods against each other, as well as against well-tuned traditional IR methods, which led to concerns~\citep{Lin_SIGIRForum2018} in the IR community as to whether ``real progress'' was being made.
Subsequently after the release of the MS MARCO dataset, some of these neural methods (\eg, \citep{dai2018convolutional, mitra2017learning}) reproduced their claimed improvements over traditional methods on the public leaderboard.
BERT put any remaining concerns to rest, as can be seen by not only the initial big jump in effectiveness as well as the continued upward progress in SOTA, in both the document and passage ranking leaderboards.
The effectiveness of BERT was widely reproduced and shown to be a robust finding, leading \citet{lin2019neural} to later retract their criticisms.

\begin{figure}[t]
\centering
\begin{tikzpicture}[scale = 1.0]
\begin{axis}[
width=0.90\columnwidth,
height=0.75\columnwidth,
legend cell align=left,
mark options={mark size=3},
font=\scriptsize,
axis y line*=left,
xmode=log,
xmin=1, xmax=530,domain=1:10,
ymin=0.05, ymax=0.40,
log ticks with fixed point,
xtick={0.5, 1, 2.5, 10, 100, 530},
ytick={0.1, 0.15, 0.2, 0.25, 0.3, 0.35, 0.4},
legend pos=south east,
xmajorgrids=true,
ymajorgrids=true,
xlabel style={font = \small, yshift=1ex},
xlabel=\# relevant query-doc training instances (thousands),
ylabel= MRR@10,
ylabel style={font = \small, yshift=0ex}]

\addplot+[
  black, mark=triangle, orange, mark options={scale=1},
  error bars/.cd, 
    y fixed,
    y dir=both, 
    y explicit
] table [x=x, y=y,y error=error, col sep=comma] {
    x,    y,       error
    
    0.5, 0.158,   0.033
    1, 0.127,   0.058
    2.5, 0.172,   0.031
    10, 0.201,  0.012
    530, 0.355, 0.000
};
\addlegendentry{\textsc{BERT-base}}

\addplot+[
  black, very thick, dashed, mark options={black, scale=0.0},
] table [x=x, y=y,y, col sep=comma] {
    x,    y
    
    0.5, 0.184
    530, 0.184
};
\addlegendentry{\textsc{BM25}}

\end{axis}
\node[above, font=\small] at (current bounding box.north) {MS MARCO Dev Set};
\end{tikzpicture}
\caption{Effectiveness of BERT-base trained with different numbers of training instances (note the log scale in the $x$-axis). Results report means and 95\% confidence intervals over five trials. Taken from~\citet{nogueira-etal-2020-document}.}
\label{figure:training_examples}
\end{figure}
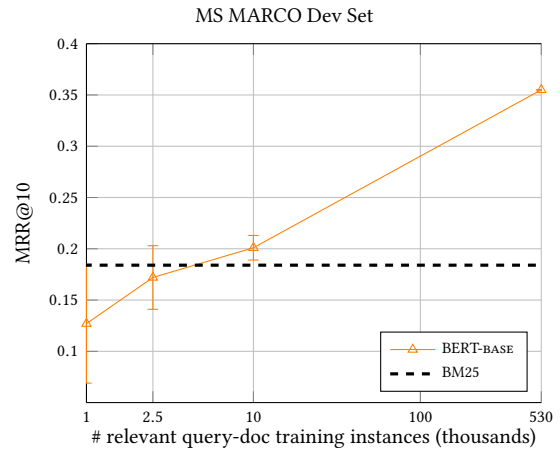

The MS MARCO datasets have been instrumental in driving this progress because it enabled all researchers (not only those in industry) to examine neural techniques in the large-data regime.
The impact of data is shown in Figure~\ref{figure:training_examples}, taken from~\citet{nogueira-etal-2020-document}.
The figure shows the effectiveness of BERT-base as a reranker trained with different numbers of training instances (note the log scale in the $x$-axis).
Results report means and 95\% confidence intervals over five trials.
As expected, the more the data, the better the effectiveness.
As pointed out by some researchers \citep{lin2020pretrained}, to a large extent, the rapid progress made in the IR community would not have been possible without MS MARCO.


So what is the state of the field at present?
We can summarize as follows:\ (1) the MS MARCO datasets have enabled large-data exploration of neural models, and (2) from the leaderboards, it appears that progress continues unabated.

But is the ``SOTA'' progress meaningful? Is MRR a good metric? Are all the top runs tied, with an exhausted leaderboard? Have we seen multiple submission and overfitting? If we change the test data slightly, as a test of external validity, do our findings hold up? Are these easy to deply, with a standard playbook? We describe what is required to make more progress, towards having many evaluation with internal validity, external validity and robust usefulness.

\section{Requirements to advance the state of the art}

This section outlines some steps that are required to make a valid and useful contribution to the state of the art in ad hoc ranking. Valid because we are sure it is an improvement. Useful because the improvement is easy to deploy in many different real-world applications. We first describe an older improvement, where significantly better rankers such as BM25 were developed using TREC data in the 1990s. We then consider the same criteria for BERT-style rankers using the MS MARCO and TREC Deep Learning Track data. This is a checkpoint on our progress so far, it motivates some of our analysis in this paper and identifies important future work.

\subsection{BM25 and TREC data}

New data can move the field forward. For example, TREC \citep{voorhees2005trec} introduced test collections starting in 1991 led to a new generation of ranking functions. The test collections did not have a large set of training queries, encouraging the development of ranking functions that work well in a small training data regime. The number of query topics used in each evaluation was 50. Compared to previous evaluation efforts, TREC documents were longer, and they varied in length, writing style, level of editing and vocabulary \citep{harman1993overview}. By the third year of the effort, this led to the development of new ranking functions that dealt with variation in document length significantly better than previous ranking functions, including Okapi BM25 \citep{robertson1995okapi}.

Today the “Okapi at TREC-3” paper has 2,420 citations in Google Scholar and searching for that string seems to mostly give papers about information retrieval (checking a few) with an estimated 15,700 results. BM25 was developed just before the appearance of the first Web search engines, but was found to work well on Web documents and was also commonly used in learning to rank data sets, many of which used web data \citep{liu2011learning}. Papers might use BM25 features in a learning to rank data set without mentioning BM25, but it still had impact. Many real-world information retrieval systems implement BM25 and it has most likely been evaluated on many proprietary data sets, not just with TREC-style evaluation, but also with online tests such as interleaving and A/B tests \citep{hofmann2016online}.

\paragraph{Internal validity} With each study, there is a risk that the conclusions we draw are not reliable. Here we focus on statistical and mathematical correctness of the study \citep{carterette2015sequential}. A study can be under powered, meaning that we can not draw finer-grained conclusions. For example, we can identify that BM25 is significantly better than a plain tf-idf implementation, but it may be statistically indistinguishable from other modern BM25-like functions.

Multiple testing and selective publication can harm the internal validity of our studies~\citep{craswell2021trec}. Statistical significance tests tell us how likely our findings may hold up on a new sample of data from the same distribution. However, if we run multiple tests on the same sample, and selectively report the best outcomes on that sample (and the bad outcomes may be rejected even if reported in a paper), the chances of that result holding on a new sample are reduced.

The best practice for avoiding multiple testing is to avoid reuse of the data, such as an online A/B test, where each new test is on live data, without reuse. Submitting to evaluation efforts such as TREC also avoids reuse, since each year generates a new set of single-shot submissions on a new set of queries. Public leaderboards are a bit worse, since they allow multiple submission to the same dataset. We will discuss methods of reducing the harm and we will analyze the extent of the problem in MS MARCO leaderboards. The most harmful case for multiple testing is with reusable test collections, which allow unlimited iteration on a test set with no public registration of what was done. There have been some claims that the field has a problem with this kind of validity \citep{armstrong2009improvements} although that paper was not questioning that BM25 was an improvement, but rather questioning whether subsequent studies improved on methods such as BM25 from the 1990s. 

If IR metrics are not on an interval scale, as was argued recently by \citet{ferrante2021towards, Fuhr2020Keynote}, this is also an internal validity problem. If commonly-used metrics such as Mean Average Precision (MAP) and Normalized Discounted Cumulative Gain (NDCG) are not on an interval scale, then reporting the mean of the metric and doing a statistical test on difference of means is not valid. Many forms of evaluation used on BM25 did calculate mean metrics with t-tests. However, a model that has been very widely deployed such as BM25 has also been tested in online interleaving and A/B tests with large numbers of users, which may not have the same problems. There is also evidence that sufficiently powered online experiments of this sort can agree with a TREC-style NDCG metric \citep{radlinski2010comparing}.

\paragraph{External validity} A study could be internally valid, with statistical tests that indicate how well the results will hold up on a new identically-distributed sample of data, but still lack external validity. Here we focus on slight changes in the data distribution, such as moving to a slightly different document distribution, query distribution or relevance judging scheme. \citet{zobel1998exploring} evaluated many BM25-style rankers on six different data distributions (which they called domains), coming from two different document collections, each with title, narrative or full queries. Their finding was that there was no clear best method with ``success in one domain was a poor predictor for success in another’’. 

The TREC finding from the 1990s, that BM25-like rankers improved on pre-TREC rankers, has good evidence of external validity. BM25 has been tested on many datasets in industry and academia, on public and private datasets, with TREC-style evaluation and presumably with online metrics. It has been selected as a powerful feature many times, by many different machine learned rankers, on many different data distributions. It would be incorrect to say that the improved performance identified in TREC-3 only held up on TREC-3 data or datasets with identical distributions.

\paragraph{Robust usefulness} BM25 is not only valuable on many different settings, but it is useful, robust and easy to deploy. It has a small number of free parameters, but if these are not tuned then the performance is still good. BM25 can be included in an IR system without needing extra training data, without needing a PhD (or PhD student) to carry out finetuning. The chances of BM25 giving very bad results in a new setting are low.

\subsection{BERT-style rankers and MS MARCO data}

In the case of MS MARCO, the main difference from TREC data is the presence of large training data, with hundreds of thousands of training queries. This encourages the development of rankers that can work well in the large-data regime, such as BERT-style rankers. Have these rankers been evaluated with internal and external validity, in a way that is robustly useful when deployed? Let us assess how far we are from this goal.

\paragraph{Internal validity} Multiple testing is a problem in our field, we discourage multiple submission in several ways. We have experiments in the TREC Deep Learning Track, where there is a single-shot submission each year, which is the gold standard for avoiding data reuse. We then retire the data as a reusable test collection, which is the worst case here, very vulnerable to multiple testing and the tests that do not show a gain may not be written up as papers and/or may not be accepted. We also have a leaderboard, which allows multiple submission, but we discourage multiple submissions. First, we limit how frequently each group submits. Second, every submission is public, so we can see which groups seem to be p-hacking and slowly overfitting to the test data through multiple submission. Third, with each submission we have a small number of queries that are not used for the leaderboard metric. We will analyze the extent to which the evaluation on these held out queries diverges from the evaluation using the queries in the leaderboard, which could happen if participants are iterating on their submissions and using the numbers on the public leaderboard as their guide.

The other threat to internal validity is whether we can find repeatable and valid differences between leaderboard runs. Perhaps the top runs are all statistically indistinguishable, and after a while we should stop the evaluation. Perhaps due to the questions in the field about the interval scale, we shouldn’t be using the mean and t-test approach that many papers in the field use. We will analyze the reliability of our leaderboard under different statistical tests and also use bootstrapping to analyze its reliability.

\paragraph{External validity} Eventually, if BERT-style rankers are widely adopted, they will be evaluated in many different settings using many different metrics. However, when we first saw good leaderboard results from ML-heavy approaches, we were suspicious that the improvements would only hold true if the training and test data were independent identically distributed (IID) samples form the same distribution. For example, there could be quirks of the MS MARCO sparse labeling that pretrained transformer models can learn, giving good performance on MS MARCO sparse labels in the test set, but the improvements would vanish if we relabeled the data with slightly different judging scheme. In that case, the results would be specific to the setup of our study, lacking external validity. We could only claim a real improvement if we think real users have exactly the same quirks as the MS MARCO labels.

We test this two ways. Firstly, we set up the TREC experiment with a slight data mismatch between the train and test data. Specifically, NIST judging selects queries that have the right level of difficulty (not too easy nor too hard) and instead of roughly one positive result per query as in MS MARCO, TREC judges label many documents per query on a 4-point relevance scale. In the DL track, we found that training on the sparse labels does allow a big improvement on the test set, despite the slightly different data distributions~\citep{trec2019overview,trec2020overview}. Second, in the document ranking leaderboard of MS MARCO we included some queries that are not used in the public leaderboard. This allows us to do a private leaderboard analysis, in this case on the 45 TREC 2020 queries, using NIST labels (as well as the sparse labels on the same queries).
These are small steps to ensure that the BERT-style rankers will perform well in many applications, these get confirmed over time in industry and academia with tests on many proprietary and public datasets.

\paragraph{Robust usefulness} We survey all the different ways people are using BERT-style rankers. We discuss our concerns about whether we really established a playbook yet, making it easy for a non-PhD to deploy this kind of ranker in a new application in a way that truly works better than previous rankers.

\section{MS MARCO leaderboard validity analysis}

To test the validity of our leaderboard, we first analyze its ability to distinguish different runs using a variety of parametric and non-parametric statistical tests. We also use bootstrapping to analyze the leaderboard stability, which in some cases can indicate that the top-ranked result was lucky (Table 4 of \citep{caruana2004kdd}). We also use bootstrapping analysis to test external validity, using our private leaderboard. These 45 TREC-2020 queries were part of every submitted run, but did not contribute to the public leaderboard numbers. We analyze whether the leaderboard conclusions generalize to these held-out queries, using sparse MS MARCO labels and also using TREC labels. Finally, since we are concerned about multiple submission, we analyze the leaderboard with respect to multiple submissions from the same group, to see if they seem to be benefitting from these submissions and whether their movement on the private leaderboard is different from that on the public leaderboard.

\subsection{Public leaderboard stability}

\begin{table}[]
    \centering
    \caption{Passage ranking leaderboard bootstrap analysis.}
    \begin{tabular}{rrrrrr}
\toprule
& \multicolumn{5}{c}{Rank under bootstrapping} \\
\cmidrule(rl){2-6}
Leaderboard run & \multicolumn{1}{c}{1} & \multicolumn{1}{c}{2} & \multicolumn{1}{c}{3} & \multicolumn{1}{c}{4} & \multicolumn{1}{c}{5}\\
\midrule
1\textsuperscript{st} & 72.7\% & 25.4\% & 1.9\% & 0\% & 0\% \\
2\textsuperscript{nd} & 24.2\% & 62.5\% & 13.3\% & 0\% & 0\% \\
3\textsuperscript{rd} & 3.1\% & 12.1\% & 83.9\% & 0.8\% & 0.1\% \\
4\textsuperscript{th} & 0\% & 0\% & 0.6\% & 47.0\% & 27.1\% \\
5\textsuperscript{th} & 0\% & 0\% & 0.2\% & 34.5\% & 34.0\% \\
\bottomrule
    \end{tabular}
    \label{tab:passage_overall_bootstrap}
\end{table}

\begin{table}[]
    \centering
    \caption{Document ranking leaderboard bootstrap analysis.}
    \begin{tabular}{rrrrrr}
\toprule
& \multicolumn{5}{c}{Rank under bootstrapping} \\
\cmidrule(rl){2-6}
Leaderboard run & \multicolumn{1}{c}{1} & \multicolumn{1}{c}{2} & \multicolumn{1}{c}{3} & \multicolumn{1}{c}{4} & \multicolumn{1}{c}{5}\\
\midrule
1\textsuperscript{st} & 91.2\% & 7.4\% & 1.4\% & 0\% & 0\% \\
2\textsuperscript{nd} & 6.8\% & 61.7\% & 21.1\% & 8.6\% & 1.4\% \\
3\textsuperscript{rd} & 1.6\% & 22.7\% & 36.8\% & 20.2\% & 12.2\% \\
4\textsuperscript{th} & 0.4\% & 5.4\% & 17.7\% & 27.0\% & 25.1\% \\
5\textsuperscript{th} & 0\% & 0.5\% & 15.9\% & 21.2\% & 22.9\% \\
\bottomrule
    \end{tabular}
    \label{tab:document_overall_bootstrap}
\end{table}

We analyze overall leaderboard stability using bootstrapping, similar to previous work by \citet{caruana2004kdd}, which avoids running many pairwise statistical tests. For each leaderboard we run 1000 bootstrapping trials, comparing the top-ranked runs, the most recent runs and baseline runs. Each bootstrapping trial samples a queryset of the same size as the original queryset, with replacement.

Our first question is whether the leaderboard's top ranks are stable under bootstrapping. If we saw that many top runs had similar chance of being top-ranked, we might conclude that the leaderboard is exhausted. It is even possible to find that the top run on the leaderboard was lucky, and some other run has more appearances at the top under bootstrapping~\citep{caruana2004kdd}.
Tables~\ref{tab:passage_overall_bootstrap} and \ref{tab:document_overall_bootstrap} show the top-5 stability of the passage and document leaderboards, respectively. It is not the case that the top ranks are all tied. The 1\textsuperscript{st} ranked run on each leaderboard never drops below position 3 in any of the 1000 trials. The tables show some indication that lower ranked reuslts are less certain, for example the 5\textsuperscript{th} ranked run has less than 50\% chance of being in position five. This can happen when two runs are similar. In the document leaderboard, the 5\textsuperscript{th} and 6\textsuperscript{th} run have expected ranks of 5.1 and 5.4 over 1000 trials.
\begin{figure}
    \centering
    \includegraphics[width=0.8\columnwidth]{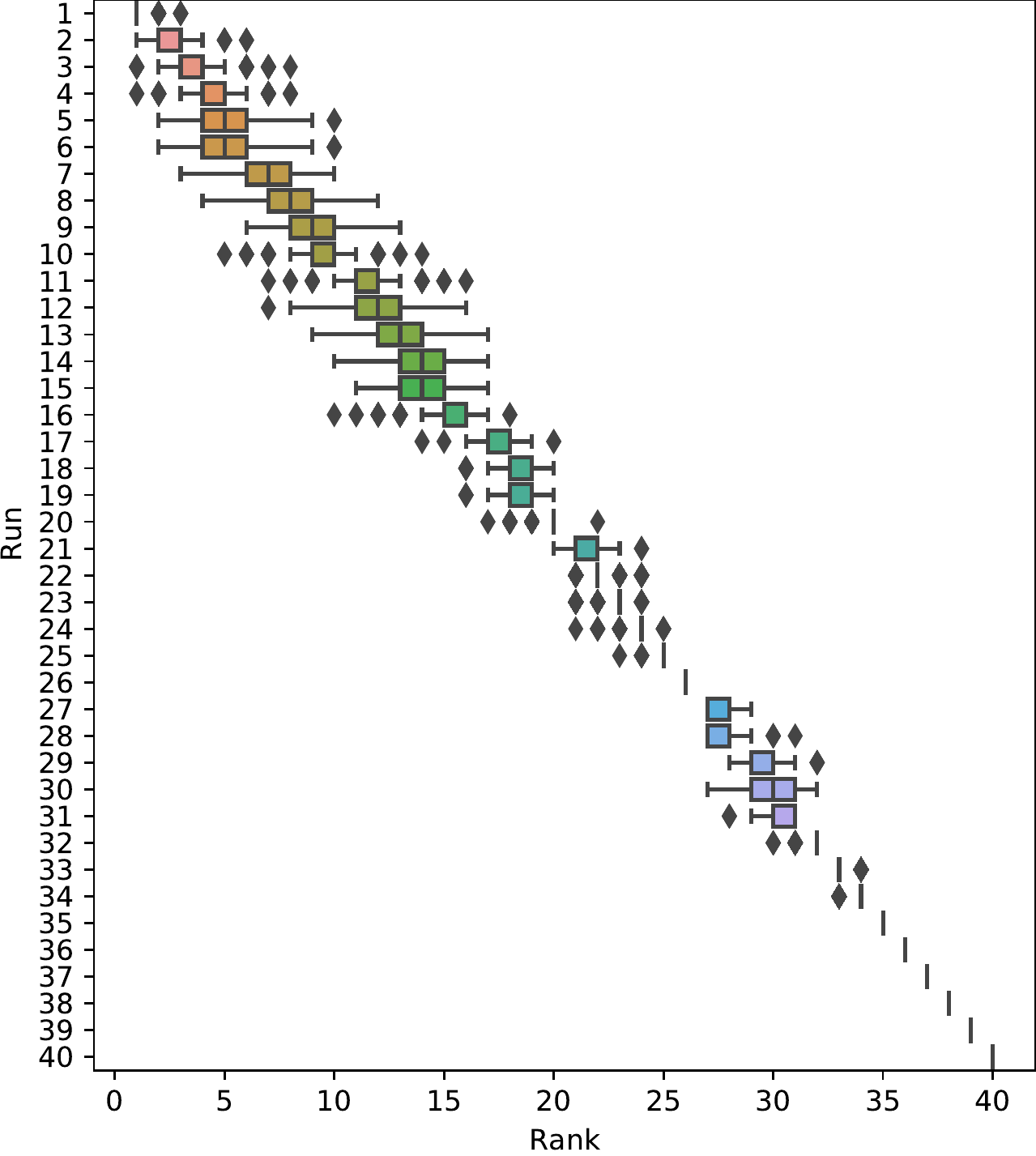}
    \caption{Full results of document leaderboard bootstrap. Runs 1--5 show the same results as Table~\ref{tab:document_overall_bootstrap}.}
    \label{fig:bootstrap_full_doc_leaderboard}
\end{figure}

\begin{figure}
    \centering
    \includegraphics[width=0.8\columnwidth]{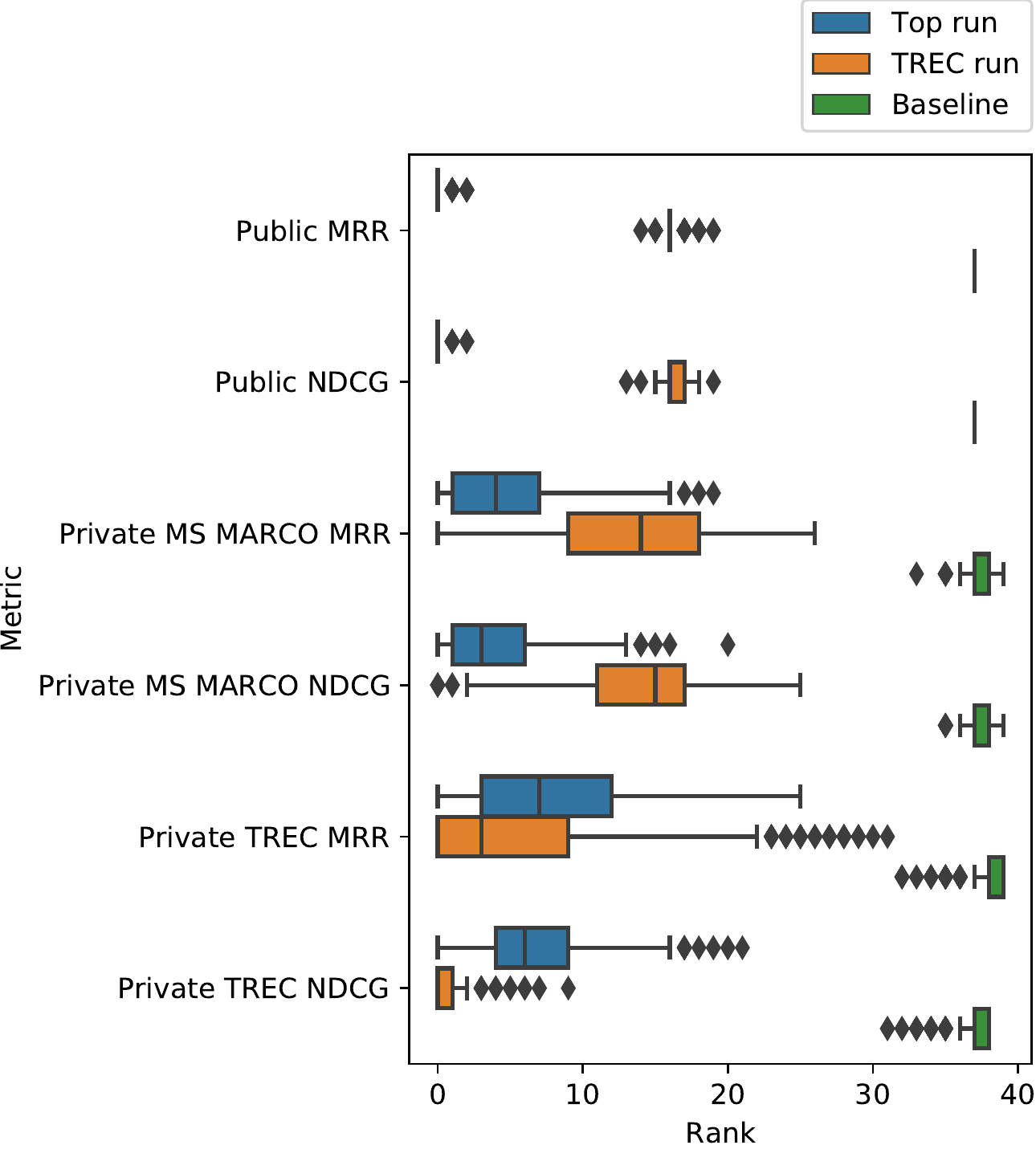}
    \caption{Rank positions of three leaderboard runs under bootstrapping. Metrics are MRR and NDCG@10. The querysets are the 5,793 Public leaderboard queries and the 45 Private leaderboard queries from TREC-2020. The Private queries can be evaluated with sparse MS MARCO labels or comprehensive TREC labels.}
    \label{fig:bootstrap_boxplot}
\end{figure}

The overall stability of the document leaderboard under bootstrapping is shown in Figure~\ref{fig:bootstrap_full_doc_leaderboard}. There are some runs with similar performance lower down in the ranking, having very similar rank distributions. The official baseline at 38\textsuperscript{th} position was ranked 38\textsuperscript{th} in all 1000 bootstrapping trials. Overall it was very unlikely under bootstrapping that a lower-ranked run would overtake a top-ranked run, leading us to conclude that the leaderboard is quite stable. The top-ranked run is not there by luck.

\subsection{Private leaderboard}

It is possible for a leaderboard to be stable, as in our bootstrapping analysis, but still have overfitting due to multiple submission. One way of detecting this is to have a private leaderboard, where each of the submissions can be tested on a held-out dataset. If participants are using the public leaderboard to overfit to the test queries, we would see their performance increase on the public query set, and decrease on the private query set.

To allow this sort of testing, we included some additional queries that were run by every participant in the document leaderboard. Here we use the 45 TREC 2020 queries for our analysis. Including our earlier bootstrapping on the Public leaderboard, we now have full bootstrapping analysis with 1000 trials on six alternatives: Metric is MRR or NDCG@10, query set is Public or Private, and relevance labels on the Private queries are sparse MS MARCO labels or comprehensive TREC labels. 

Instead of showing the full bootstrapping results for all six combinations, we summarize three key runs and their performance under bootstrapping. Figure~\ref{fig:bootstrap_boxplot} shows this analysis. The top run on the public leaderboard has its ranks more spread out on the smaller Private leaderboard queries, since there are 45 rather than 5793 queries. It is overtaken by other runs not only on certain bootstrap trials, but other runs even have a better expected rank. For the MS MARCO labels, the top leaderboard run is ranked fourth in expectation for MRR and third in expectation for NDCG@10. For the TREC labels, the top run is fifth and sixth.

To explain why we saw greater rank for the TREC labels, we note that some runs submitted to the leaderboard were also submitted to TREC and may have used the TREC 2019 labels for training. The TREC run we highlight in the figure is the ranker from University of Waterloo, that achieved the best NDCG@10 at TREC. This could be seen as a lack of external validity, that the top run on MS MARCO labels is not as highly-ranked on TREC labels. It could also be seen as an indication that extra adaptation and training for the target domain is useful. Overall the official baseline run is significantly worse than our top run and TREC run, under all six conditions. 

\subsection{Multiple submission}

To avoid overfitting to our eval queries, the MS MARCO leaderboards have rules limiting multiple submission. We allow each participating group to submit no more than two runs per month, and no more than one run with very small changes such as hyperparameter tuning or random seeds. This makes it more difficult for participants to try minor variations until they get lucky with a higher leaderboard submission. We also track all the submissions, so if a group is submitting many runs we can analyze what they submitted and how often. We believe this makes it much more difficult for participants to overfit, compared to a reusable test collection which allows unlimited iteration with no public record.

We already presented detailed bootstrapping results for the document leaderboard. Now, we consider the institution that submitted the top run, and whether there is a risk of overfitting. On the document leaderboard, grouped by institution, the three institutions with the most submissions had 12, 11 and 7 submissions. However, the top run came from an institution with only two submissions. Also, on Private MS MARCO evaluation (Figure~\ref{fig:bootstrap_boxplot}) the top run is still in the top five runs in expectation. If it were only there by overfitting, we do not think it would be in the top five of forty.

There is still a risk of cross-group overfitting, as groups learn about ``what works'' some of what they learn may be about this samlple of test queries. For example, different groups can share code and ideas. They can converge to a common solution, with many groups submitting slight variations, and one group can get lucky. Through this process of code sharing they can also form an ensemble of promising approaches, and since the promising approaches were selected using the evaluation set, this is another form of overfitting. In future we will use the MS MARCO judgments on the private leaderboard to monitor for overfitting.

\section{IR metrics and the interval scale}

Commonly used information retrieval metrics, including the ones we employed in our leaderboard evaluation such as NDCG~\citep{JK2002} and MRR~\citep{craswell2009mean}, have recently been criticized by \citet{ferrante2017ir, ferrante2018general} for not being interval-scale, which would imply that computing their mean values across different queries is not meaningful.
Instead, they argue that most IR metrics tend to be in ordinal scale, implying that we should be using the median metric value as opposed to the mean when we aggregate these values across different queries.
This ignited a debate in the IR community with \citet{fuhr2018some} arguing that it is, therefore, not meaningful to compute the mean of MRR and ERR metrics over multiple relevance topics.
\citet{sakai2020fuhr} subsequently disagreed citing that this line of reasoning would render many other IR metrics inappropriate and that many of these metrics are practically useful, even if not theoretically justified.
More recently, \citet{ferrante2021towards} have furthered the argument made by \citet{fuhr2018some} to point out that indeed for many well-known and commonly used IR metrics it is inappropriate to compute their average.

Because the MS MARCO labels are binary and sparse, we chose to report MRR as our primary metric on the leaderboard.
Similarly at TREC, the Deep Learning track has focused on NDCG, NCG~\citep{rosset2018optimizing}, MRR, and MAP~\citep{zhu2004recall}.
So, the validity of these metrics is an important consideration in the context of benchmarking on MS MARCO.
Our position in this paper is that \citet{ferrante2021towards} have raised a valid issue and indeed there is no reason to assume that metrics like MRR, MAP, and NDCG are on an interval scale.
However, we do not fully agree with their theoretical argument and recommendations, and present an alternative viewpoint here.

\subsection{Preliminaries}
The theoretical argument presented by \citet{ferrante2021towards} is grounded in the \emph{representational theory of measurement}~\citep{krantz2006additive} which views measurement as the process of mapping real world entities to numbers such that some entity attributes are represented faithfully as numerical properties.
Before we analyze their argument, we define a few preliminary concepts and notations for our reader.
We adopt the same notation as \citet{ferrante2021towards} here for consistency.

\theoremstyle{definition}
\begin{definition}[Relational structure]
A \textbf{relational structure} is an ordered pair $\mathbf{A}=\langle A, R_A \rangle$ where $A$ is a domain set and $R_A$ is a set of relations on $A$.
If the $A$ is a set of entities then we refer to it as an \textbf{empirical relational structure}.
In contrast, in case of \textbf{numerical} or \textbf{symbolic relational structure} $A$ is a set of numbers.
\end{definition}

\theoremstyle{definition}
\begin{definition}[Homomorphism]
Given two relational structures $\mathbf{A}_1$ and $\mathbf{A}_2$, a homomorphism $\mathbf{M}: \mathbf{A}_1 \to \mathbf{A}_2$ is a mapping $\mathbf{M}: \langle M, M_R\rangle$ such that,
\begin{enumerate}[label=\roman*.]
    \item The function $M$ maps $A_1$ to $M(A_1) \subseteq A_2$
    \item The function $M_R$ maps $R_{A_1}$ to $M(R_{A_1}) \subseteq R_{A_2}$, such that $\forall r \in R_{A_1}$, $r$ and $M(R_{A_1})$ have the same arity
    \item $\forall r \in R_{A_1}$, if the relation $r$ holds between some elements from the domain set $A_1$ then the image relation $M(R_{A_1})$ should also hold for the corresponding image elements in $A_2$.
\end{enumerate}
Note that we use homomorphism instead of isomorphism because $M$ is typically not a one-to-one mapping.
\end{definition}

\theoremstyle{definition}
\begin{definition}[Measurement]
A \textbf{measurement (scale)} is the homomorphism $\mathbf{M}: \mathbf{E} \to \mathbf{N}$ that maps from the empricial relation structure $E$ to the numerical relational structure $N$.
The mapping of an element $e \in E$ to a number $n \in N$ is called a \textbf{measure}.
\end{definition}

\theoremstyle{definition}
\begin{definition}[Difference structure]
An empirical relational structure $\mathbf{E}=\langle E, \preceq \rangle$ is a \textbf{difference structure} if $\forall a,b \in E$ it defines a \textbf{difference} $\Delta_{ab}$ and satisfies the following axioms:
\begin{enumerate}[label=\roman*.]
    \item $\preceq$ is a weak order---\ie, $\preceq$ is a binary relation on $E \times E$ such that $\forall a,b,c \in E$ it satisfies:
    \begin{enumerate*}[label=(\alph*)]
        \item $a \preceq b$ or $b \preceq a$, and
        \item $a \preceq b$ and $b \preceq c$ $\implies a \preceq c$.
    \end{enumerate*}
    \item $\forall a,b,c,d \in E$, $\Delta_{ab} \preceq \Delta_{cd} \implies \Delta_{dc} \preceq \Delta_{ba}$
    \item $\forall a_1,b_1,c_1,a_2,b_2,c_2 \in E$, $\Delta_{a_1b_1} \preceq \Delta_{a_2b_2}$ and $\Delta_{b_1c_1} \preceq \Delta_{b_2c_2} \implies \Delta_{a_1c_1} \preceq \Delta_{a_2c_2}$
    \item $\forall a,b,c,d \in E$, if $\Delta_{aa} \preceq \Delta_{cd} \preceq \Delta_{ab}$, then there exists $x,y \in E$ such that $\Delta_{ax} \sim \Delta_{cd} \sim \Delta_{yb}$ (Solvability Condition)
\end{enumerate}
\end{definition}

According to the \emph{representation theorem} for difference structures, if there is a difference structure on the empirical set $E$ then there must exist an interval scale $M$.

\subsection{Analysis of argument by Ferrante et al.}
Having covered the preliminaries, we now take a closer look at the argument that \citet{ferrante2021towards} make in their work.
They define a domain set over search result page (SERP) states, where each SERP state is a unique rank-ordered list of relevance grades.
For example, under this notation a SERP with three documents with a relevant document at rank two and nonrelevant documents at rank one and three corresponds to a SERP state denoted by the tuple $(0, 1, 0)$.
For example, if we consider the universe of all SERPs with exactly three documents and binary relevance grades, then the domain set $E$ over all possible SERP states is $S = \{(1,1,1), \allowbreak (1,1,0), \allowbreak (1,0,1), \allowbreak (1,0,0), \allowbreak (0,1,1), \allowbreak (0,1,0), \allowbreak (0,0,1), \allowbreak (0,0,0)\}$.
\citet{ferrante2021towards} argue that if we can define a difference structure over $\mathbf{S}=\langle S, \preceq \rangle$ then it would imply the existence of a corresponding interval scale.
However, for $\mathbf{S}$ to satisfy the Solvability Condition requires the metric to be equi-spaced between any two neighboring items in a partial ordering of the domain set $S$.
For example, Table~\ref{tbl:example} shows that MRR can take four discrete values $[1.00, 0.50, 0.33, 0.00]$ in context of the same example scenario with SERPs of fixed length three and binary relevance grades.
If we consider the four specific SERP states labeled A--D, we observe that the Solvability Condition is violated because the presence of $\Delta_{CD}=0.17$ implies there should exist $X,Y \in S$, such that $\Delta_{AX}=\Delta_{YB}=\Delta_{CD}=0.17$.
The key argument that \citet{ferrante2021towards} make is that MRR is not interval-scale because values of $0.17$ and $0.83$ are not realizable under this setting.


\begin{table}
    \setlength\tabcolsep{3.5pt}
    \small
    \centering
    \caption{A tabular representation of the domain set $S$ of all SERPs with exactly three results with binary relevance grades and corresponding MRR values.}
    \begin{tabular}{l|llllllll}
    \hline
        & A & & & & C & & D & B \\
        \textbf{$s \in S$} & (1,1,1) & (1,1,0) & (1,0,1) & (1,0,0) & (0,1,1) & (0,1,0) & (0,0,1) & (0,0,0) \\
        \textbf{RR} & $1.00$ & $1.00$ & $1.00$ & $1.00$ & $0.50$ & $0.50$ & $0.33$ & $0.00$ \\
        \hline
    \end{tabular}
    \label{tbl:example}
\end{table}

However, our position is that relevance metrics like MRR and NDCG are fundamentally not measurements over SERP states, but instead they measure user perceived relevance of the SERPs.
Hence, the difference structure should not be applied on the domain set $S$ of all possible SERP states, but instead on the domain set $U$ of all possible user-perceived relevance states.
We argue that an appropriate IR metric should be equi-spaced relative to user perception of relevance such that a change in $0.1$ in the metric at any point on the scale (\eg, $0.3 \to 0.4$ \vs $0.75 \to 0.85$) should correspond to same difference in user-perceived relevance.
In other words, it is irrelevant if a three-document SERP cannot realize a MRR value of $0.17$ as long as we believe that there exists some user-perceived relevance state that corresponds to that value of the metric.

Now, there is no reason to believe that IR metrics without further calibration would be equi-spaced on the scale of user-perceived relevance.
We therefore agree with \citet{ferrante2021towards} that computing mean of many IR metrics may be inappropriate.
But the difference between their argument and ours points to different recommendations for addressing these concerns.
To remedy the situation, \citet{ferrante2021towards} propose ranked versions of common IR metrics that are equi-spaced over $E$.
While the mean value of these ranked-metrics may be more meaningful from the viewpoint of \emph{representational theory of measurement}~\citep{krantz2006additive}, it is possible, if not likely, that it reduces the correspondence of the metric to user-perceived relevance. By our criteria, the better approach is to conduct lab studies and online studies with real users, to understand their preferences and how this reveals their underlying notion of utility, so we can develop metrics that are on an interval scale in user value. 





\subsection{Reliability of statistical tests}

\begin{table}
    \setlength\tabcolsep{1pt}
    \small
    \centering
    \caption{Agreement rates for different significance tests across 100 different query set splits for different task and metric combinations.}
    \begin{subtable}{\columnwidth}
    \centering
    \caption{document ranking using MRR}
    \begin{tabular}{ |c|c|c|c|c|c|c|c| } 
\toprule
 & Sign T. & WX RS & WX SR & t-test & Sign T. Med. &	WX RS Med & WX SR Med \\
 \midrule
 agree & 93.3\% & 92.1\% & 91.2\% & 91.7\% & 80.2\% & 81.2\% & 80.7\% \\
 part. agree & 3\% & 3\% & 3\% & 3\% & 16.1\% & 16.1\% & 16.1\% \\ 
 disagree &  3.7\% & 4.9\% & 5.8\% & 5.3\% & 3.7\% & 2.7\% & 3.2\% \\ \hline
 perc. signif.& 95.0\% & 79.7\% & 92.7\% & 79.8\% & 95.0\% & 79.7\% & 92.7\% \\
\bottomrule
    \end{tabular}
    \label{tab:document_significance_mrr}
    \end{subtable}
    \begin{subtable}{\columnwidth}
    \centering
    \caption{passage ranking using MRR}
    \begin{tabular}{ |c|c|c|c|c|c|c|c| } 
\toprule
 & Sign T. & WX RS & WX SR & t-test & Sign T. Med. &	WX RS Med & WX SR Med \\
 \midrule
 agree & 92.8\% & 91.6\% & 90.8\% & 90.8\% & 80.6\% & 81.4\% & 81.2\% \\
 part. agree & 3.3\% & 3.3\% & 3.3\% & 3.3\% & 15.5\% & 15.5\% & 15.5\%\\ 
 disagree &  3.9\% & 5.1\% & 5.9\% & 5.9\% & 3.9\% & 3.1\% & 3.3\%\\ \hline
 perc. signif.& 94.6\% & 79.8\% & 92.8\% & 79.9\% & 94.6\% & 79.8\% & 92.8\%  \\
\bottomrule
    \end{tabular}
    \label{tab:passage_significance_mrr}
    \end{subtable}
    \begin{subtable}{\columnwidth}
    \centering
    \caption{document ranking using NDCG}
    \centering
    \begin{tabular}{ |c|c|c|c|c|c|c|c| } 
\toprule
 & Sign T. & WX RS & WX SR & t-test & Sign T. Med. &	WX RS Med & WX SR Med \\
 \midrule
 agree &    94.9\% & 92.2\% & 91.0\% & 91.8\% & 85.2\% & 84.2\% & 83.3\% \\
 part. agree & 1.8\% & 7.7 \% & 8.4 \% & 8.1\% & 1.8\% & 10.4\% & 6.2\% \\ 
 disagree   & 3.2\% & 0.1\% & 0.6\% & 0.1\% & 13.0& 5.4\% & 10.5\% \\ \hline
 perc. signif.& 97.9\% & 81.1\% & 93.3\% & 81.4\% & 97.9\% & 81.1\% & 93.3\% \\
\bottomrule
    \end{tabular}
    \label{tab:document_significance_ndcg}
    \end{subtable}
\end{table}


In this section we analyse the effect of IR metrics not being interval-scale on evaluation outcomes in practice. Apart from the mean not being very meaningful when aggregating metrics that are not in interval scale across different queries, \citet{ferrante2021towards} have also raised concerns about the reliability of using some of the commonly used significance tests such as t-test or the Wilcoxon Signed Rank, which require that the values are in interval scale. They have instead argued that sign test or the Wilcoxon Rank Sum test should be used with ordinal measurements, such as most top heavy IR metrics. 

Aforementioned issues raised by \citet{ferrante2021towards} could also raise questions regarding the reliability of the evaluation results obtained through leaderboards like MS MARCO. 
Previous work~\citep{hull1993using, sanderson2005information} has indicated that violation of certain assumptions by some significance tests, in particular, the normality assumption for the t-test, do not have a big effect on the conclusions reached using such tests in practice. This raises the question as to how much effect the metric not being in an interval scale could affect the reliability of evaluation results obtained using different significance tests, or using different aggregation methods (i.e., mean vs. median). In order to answer this question, we adopt a similar method as the one used by Buckley and Voorhees~\cite{Buckley00} for evaluating evaluation stability. 

We divide our query set into two random subsets and for each pair of systems submitted to the leaderboard, we evaluate as to whether the conclusions reached based on the evaluation results obtained using the two subsets agree with each other. We repeat this process 100 times, generating 100 random splits and compute the agreement rates across the different subsets. If the evaluation results are reliable, we expect the results to be robust to the changes in the query sample and hence, the agreement rates should be high.

When we compare evaluation results across the two subsets, we use the following definition for agreement, partial agreement and disagreement. Evaluation results in the two subsets: 

\begin{itemize}
    \item \textbf{Agree} with each other if the two subsets agree as to which system is better, and the difference is:
    \begin{enumerate*}[label=(\roman*)]
        \item statistically significant according to both subsets, or
        \item not significantly different according to both subsets,
    \end{enumerate*}
    \item \textbf{Partially agree} with each other if
    \begin{enumerate*}[label=(\roman*)]
        \item the two subsets agree as to which system is better, and the difference is significant according the one subset but not significant according to the other, or
        \item the two subsets disagree as to which system is better, but the difference is not statistically significant according to both sides, 
    \end{enumerate*}
    \item \textbf{Disagree} with each other if the two subsets disagree as to which system is better, and the difference is:
    \begin{enumerate*}[label=(\roman*)]
        \item statistically significant according to both subsets, or
        \item statistically significant according the one subset but not significant according to the other
    \end{enumerate*}
\end{itemize}

We report the results of this experiment using MRR as the metric for the document and passage ranking tasks in Table~\ref{tab:document_significance_mrr} and~\ref{tab:passage_significance_mrr}, respectively. 
Each colum in the tables shows the agreement rates obtained when a different significance test is used in evaluation, focusing on the sign test (Sign T), Wilcoxon Rank-Sum test (WX RS), Wilcoxon Signed-Rank test (WX SR), and t-test as the significance tests. Since previous work has argued for using the median instead of the mean when the metrics are in ordinal scale, we also report the agreement rates for the significance tests that do not have the interval scale requirement (sign test and Wilcoxon Rank-Sum test), when median is used to compute the aggregate performance across different queries. While the Wilcoxon Signed-Rank test does require the metrics to be interval-scale~\citet{ferrante2021towards}, since the null hypothesis for this test is that the median (as opposed to the mean) of the differences is zero, we also report the results for this test when median is used for aggregation. The last three columns in the tables show the agreement rates when median (Med.) is used for aggregation instead of the mean.
As seen in the tables, when mean is used for aggregation, the agreement rates for all four significance tests are above 90\%. This is true both for significance tests that require interval measurements (t-test and Wilcoxon Signed-Rank test) and also for those that can be used with ordinal measurements (Sign Test and Wilcoxon Rank-Sum test). 

One potential reason for the high agreement rates could be caused by a test not being very powerful and hence mostly predicting differences as not statistically significant. Hence, we also report the fraction of pairs of systems that were deemed as significantly different by at least one of the two split sets using a particular significance test, which is reported in the last row of the tables. It can be seen that the agreement rates are not really correlated with the percentage of pairs a test identifies as significantly different. 

When median is used instead of the mean, agreement rates drop significantly and consistently across the three significance tests. Our results suggest that, even though the most commonly used IR metrics are not on interval scale, reliability of evaluation results obtained are not widely affected by this. In fact, unlike what was recommended before, using mean instead of the median seems to result in more reliable evaluation results, possibly caused by mean being a more discriminatory statistic than the median. Our results seem consistent across both document and passage ranking tasks.

In Table~\ref{tab:document_significance_ndcg} we show the results for the document ranking task when NDCG@10 is used as the evaluation metric. As expected, NDCG@10 results in higher agreement rates consistently for all significance tests when compared to MRR, even though the difference is not very big. Similar results were observed for the passage ranking task.
Overall, our results suggest that even though most commonly used IR metrics such as MRR and NDCG@10 may not be in interval scale, evaluation results obtained in practice seem not to be highly affected and results obtained using benchmarks such as MS MARCO seem to be mostly reliable. 






\section{On transfer learning from MS MARCO to other IR benchmarks}
\label{sec:transfer_learning}

The primary motivation behind curating the MS MARCO ranking datasets was to answer the question ``How much better can our IR systems be if we had access to millions of positively labeled query-document pairs?''
It is exciting to witness the large jumps in performance metrics on this benchmark from the development of new ranking models that can adequately leverage the provided large training datasets.
However, if the benefits of MS MARCO's large training data is limited to its own test sets and access to domain-specific large training datasets is only limited to large for-profit private institutions---\eg, major commercial search engines---then the creation of such benchmarks only serves to outsource research and development of models to the academic community that ironically the academic community then cannot operationalize for their own scenarios.
To avoid this undesirable dynamic, it is important to also study whether the large training dataset from MS MARCO can bring about meaningful improvements from transfer learning to other IR benchmarks and tasks.

As noted earlier, a successful application of transfer learning from the MS MARCO dataset has been for the TREC Deep Learning track.
An initial test set of 200 queries is sampled from the MS MARCO distribution, but then NIST selects a subset of queries to judge which are neither too difficult nor too easy, then apply a 4-point labeling scheme to results pooled from submitted runs.
As \citet{trec2019overview, trec2020overview} have reported, several pretraining-based deep models finetuned on the MS MARCO training data achieve significant improvements over traditional IR methods in this setting.

Transfer learning from MS MARCO to other ad hoc retrieval benchmarks have also been attempted with promising early success.
\citet{yilmaz2019cross} finetune a BERT-based~\citep{devlin2018bert} model on MS MARCO, TREC CAR~\citep{dietz2017trec} and TREC Microblog~\citep{lin2014overview} datasets and evaluate them on  three TREC newswire collections: Robust04~\citep{voorhees2004overview}, Core17~\citep{allan2017trec}, and Core18.
They find that finetuning on MS MARCO alone achieves mixed results on these benchmarks, but finetuning on MS MARCO followed by further finetuning on the TREC Microblog dataset achieves state-of-the-art performance on all three test sets.
Since then, the combination of finetuning on MS MARCO followed by on TREC Microblog dataset has also achieved state-of-the-art results on the English subtask of the NTCIR15 WWW-3 task~\citep{sakai2020overview, shindenkasys}.
Recently, \citet{nogueira2020document} adapted T5~\citep{raffel2019exploring}, a pretrained sequence-to-sequence model, by finetuning only on MS MARCO to significantly improve over the previous state-of-the-art results reported by \citet{yilmaz2019cross} on Robust04.
Similar strategies of finetuning on MS MARCO and evaluating on Robust04, GOV2~\citep{clarke2004overview}, and ClueWeb~\citep{clarke2009overview} have been employed in other recent studies~\citep{li2020parade, zhang2020little, jiang2020long, gao2020modularized, zheng2020bert}, sometimes in combination with weak supervision~\citep{sun2020meta, zhang2020selective}.
Additionally, \citet{ma2020prop} have employed the document collection in MS MARCO for pretraining before evaluating on these other standard IR benchmarks.

An interesting implication of the large size of the MS MARCO training dataset is that it allows for further filtering to generate new domain-specific training datasets that may be adequately large to finetune deep models specializing in a given domain.
This is particularly interesting when due to time sensitivity or resource constraints it is infeasible to curate a domain-specific training dataset from scratch.
Such a scenario emerged in 2020, when in response to the COVID-19 pandemic, the body of academic literature on Coronavirus grew significantly which in turn posed a difficult challenge for the information retrieval community to quickly devise better methods for searching over this growing scientific corpus.
This prompted the creation of the Covid-19 Open Research Dataset (CORD-19)~\citep{wang2020cord} and the TREC-COVID~\citep{voorhees2020trec, roberts2020trec} benchmarking effort on one hand, and a flurry of new research and development of IR systems specializing on this task~\citep{connor2021deep, chen2020comparative, wang2020text} on the other.
In particular, \citet{macavaney2020sledge, macavaney2020sledgez} created Med-MARCO, a subset of the MS MARCO dataset that are related to medical questions.
Subsequently, several groups benchmarking on TREC-COVID employed this subset for model training~\citep{zhang2020rapidly, zhang2020covidex, xiong2020cmt, liuadapting}, while others explored finetuning on the full MS MARCO for this task~\citep{sun2020meta, bendersky2020rrf102, li2020parade, nguyen2020pandemic, liuadapting, rughbeer2021dataset}.
In a meta-analysis of participating runs in the TREC-COVID challenge, \citet{chen2020comparative} found the use of MS MARCO dataset for finetuning to be associated with higher retrieval performance.
Similar to Med-MARCO~\citep{macavaney2020sledge, macavaney2020sledgez}, \citet{hamzei2019place} studies place-related subset of the MS MARCO dataset.
Another interesting case study in this context is the application of MS MARCO to conversational search where it has been useful for both creation of new benchmarks~\citep{dalton2020cast, dalton2020trec, ren2020conversations} and model training~\citep{voskarides2019ilps, kumar2020making, yu2020few, wang2019exploring, yang2019query, voskarides2020query, ferreira2021open, stamatis2019ves, mele2020topic}.
The adoption of MS MARCO in so many transfer learning settings is encouraging, and while it may be premature to draw parallels between its impact on the IR community and what ImageNet~\citep{deng2009imagenet, russakovsky2015imagenet} did for computer vision research, the current trends definitely bode well for MS MARCO's potential role in the future of IR research.

\section{Robust usefulness and externalities}

For rankers based on pretrained transformers to become a standard solution in research and industry, we need to show that they can be easily be deployed in new settings. Section~\ref{sec:transfer_learning} indicated that models can work well in a new target domain, but this may involve domain-specific data and multiple stages of finetuning. Future research could work on developing a ``play book'' for ranker deployment, with the goal of simplifying the process and decreasing the chances of problems or failure. This could include development of self-tuning rankers that can learn from the corpus and/or usage logs when deployed. It could also include the development of a general-purpose ranker, that works reasonably well in a new application with no additional finetuning.

Considering issues of deployment raises the common adage ``you can't improve what you don't measure''. Data sets and evaluation efforts have an incentive structure, that encourages work towards exactly what is measured, creating blind spots in other areas. 
MS MARCO not only serves to compare existing IR methods but also plays the Pied Piper guiding a significant section of the community down specific lanes of research.
As the curator of such benchmarks, it is therefore crucial that we critically reflect on where we are going and also importantly where we are choosing not to invest.

For example, the availability of a large training dataset directly incentivizes new methods that can take advantage of millions of labeled query-document pairs.
Excitement in the large-data work means we may see too few submissions in our benchmarking of methods in the small-data regime, even though such approaches have advantages of efficiency and robustness.
Similarly, MS MARCO is English-only, reducing our likelihood of seeing related advances in non-English and cross-language IR.

Both the MS MARCO leaderboard and the TREC Deep Learning track focuses singularly on measuring the relevance quality of retrieved documents and passages, without any consideration for other critical aspects such as efficiency or cost of deployment.
This could for example lead the community towards building new models that are frustratingly hard for others, with limited compute resources, to further optimize or deploy.
That could again create a divide between what we focus on as a community and what is practically useful.
The scenario may be even more serious if we were to consider the potential social harms---specifically on those who belong to historically marginalized communities---and ecological costs of large language models~\citep{bender2021dangers}, exactly the type of technology that MS MARCO and TREC Deep Learning track may encourage us to work on.
As we develop these new benchmarks, the responsibility rests squarely on our own shoulders to think broadly and have open and inclusive conversations about the impact of leading a large section of our community down a given path.


\section{Conclusion}

The MS MARCO leaderboards and TREC Deep Learning track have led to several new ranking approaches, and we have considered multiple aspects of the validity of such studies and usefulness of such approaches.
Our bootstrapping analysis showed that the leaderboards are quite stable, meaning that we can be highly confident that we are distinguishing between new methods and our baseline. Since a stable leaderboard can still have overfitting through multiple submission, we limit submissions per group and we have a private leaderboard that we monitor. We also have two evaluation efforts with different labeling schemes, where we use both schemes in both efforts to see if our results are robust. Given potential problems with multiple testing on reusable test collections, we agree with recent SIGIR keynotes~\citep{fuhrsigir,voorhees2020coopetition} that the gold standard is to submit to evaluation efforts such as TREC, and we also found value from having an associated leaderboard with appropriate safeguards.

Since IR metrics may not be interval-scale, and our leaderboard uses one of the most-criticized of these metrics (MRR), we analyzed our leaderboard also using NDCG, both in the bootstrapping analysis and in a test focused on the reliability of statistical tests. We found similar results using MRR and NDCG. Using a variety of statistical tests, some of which do not assume an interval scale, we found a reasonable level of reliability in results. We also argued that there is probably a big gap between true user-percieved utility and current IR metrics, so we suggest work in closing this gap.

We noted that there has been a lot of progress in adapting MS MARCO to other applications, in some cases with multi-stage finetuning. Although this is promising, it suggests that we can not simply deploy a ranker in a new domain without significant data collection and machine learning work. To truly make the new rankers robust and useful, we should make them easier to deploy. We also note a variety of blind spots in our evaluation efforts, suggesting new directions for data and evaluation efforts in the future. 

\bibliographystyle{ACM-Reference-Format}
\bibliography{bibtex}

\end{document}